# Architecture of the LHCb muon Frontend control system upgrade

Valerio Bocci, *Member, IEEE*

*Abstract*–The LHCb experiment(Fig. 1), that is presently taking data at CERN (European Center for Nuclear Research) Large Hadron Collider (LHC), aims at the study of CP violation in the B meson sector. Its key elements is the Muon detector [1], which allows triggering, and muon identification from inclusive b decays. The electronic system (Fig. 2) of the whole detector is very complex and its Muon detector Experiment Control System (ECS) allows monitoring and control of a number of Front-End boards in excess of 7000. The present system in charge of controlling Muon detector Front-End (FE) Electronics consists of 10 Crates of equipment; each crate contains two kinds of modules: a Pulse Distribution Module (PDM) and up to 20 Service Boards (SB) connected via a custom Backplane for a total amount of about 800 microcontrollers[2]. LHCb upgrade is planned for 2018/19, which will allow the detector to exploit higher luminosity running. This upgrade will allow the experiment to accumulate more luminosity to allow measurements that are more precise.

The main idea of the new architecture take advantage of the new CERN fast communication protocol GBT[3] developed for radiation environment leaving unchanged the huge connectivity to the detector and the modularity of the system. The new chipset GBTx and GBT-SCA[4] developed at CERN and the availability of more powerful computers allow designing a new system with fast link to the detector. In the new system, the control unit are moved from the apparatus to the computers in control room instead of the actual system where the low speed bus send high level commands to intelligent unit in the detector.

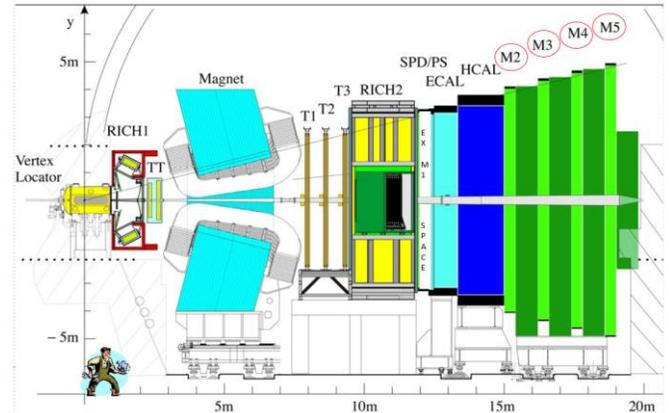

*Fig. 1 LHCb detectors M2-M5 Chambers Front End electronics upgrade.*

## I. INTRODUCTION

The LHCb Muon detector allows triggering and muon identification from inclusive b decays. This detector is based on more of 1000 high efficiency chambers, which form about 100,000 physical output channels. Read-out Electronics, composed by 7000 CARDIAC boards (CARioca DIAlog Connection) which allows forming the analog signal, take to muon chambers and generating the logical channel. The CARDIAC Front-end boards use two kinds of chip, two Carioca front-end modules containing 8 channels analog preamplifiers and discriminators, and a programmable digital chip called Dialog [5].

DIALOG integrates important tools for detector time alignment, physical channel monitor, threshold setting. In particular, it integrates 16 programmable delays, which can be regulated in steps of 1 ns, 16 counters 16 bits wide for channel rate, threshold settings.

External communication of the Front-End is based on a serial protocol, relying on LVDS physical standards, by means the LHCb Muon detector Experiment Control System (ECS) is able to monitor and control front-end electronics.

The present system in charge of controlling Muon detector Front-End (FE), Electronics is based on distributed microcontrollers (up to 810). The system consists of 10 Crates of equipment; each crate contains two kinds of modules: a Pulse Distribution Module (PDM) and up to 20 Service Boards (SB)

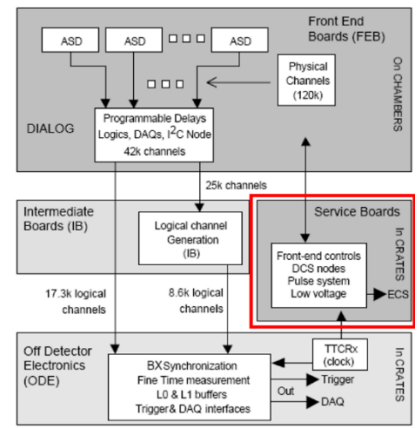

*Fig. 2 Service Board Front End Control system role.*

Valerio Bocci (e-mail: Valerio.Bocci@roma1.infn.it) INFN Sezione di Roma , P.le Aldo moro 2, Rome, I-00185, Italy.

connected via a custom Backplane, Each SB controls with 12 serial link (Fig. 3).

The new muon Frontend control system maintains the original architecture of the system for what concerns crate allocations and module partitions, cable disposition and connection.

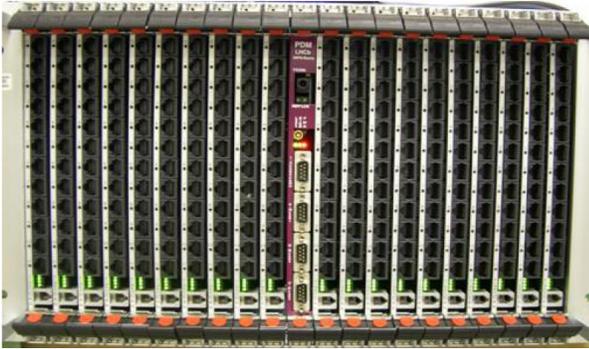

*Fig. 3 Service Board and PDM Crate without cables connection.*

## II. THE NEW PULSE DISTRIBUTION MODULE (nPDM)

In the upgraded system, a new PDM (nPDM) module (Fig. 4) is connected to main computers using an optical fiber GBT link. GBTx is a radiation tolerant chip that implements the GBT protocol. It is a high-speed bidirectional 3.2 Gb/s optical link especially designed for the communication in the LHC accelerator and experiments. The GBTx replace, with a single link, different data path communication used in the previous phase of LHC, Timing and Trigger Control (TTC), Data Acquisition (DAQ) and Slow Control (SC) information.

In the case of nPDM, the GBTx is used to receive timing information in substitution of old TTCrx chip and the Slow control command previously operated using the CANBUS communication. The GBTx bandwith can be splitted in 80 Mbit/s links called E-links in flexible way.

In order to avoid a total or partial re-cabling of more than a thousand cables between the SB system and the front-end electronics, the cabling architecture and the modularity of the system is not changed. The system, as the old one, use the same number of nSB per crate (20) and the same number of connection to the Front end per Service Board (12). The modularity of 20 Service Board per crate is maintained splitting the bandwidth of the GBTx in 80 Mbits/s per E-links using 1.6 Gbit/s of the 3.2 Gbit/s for the nPDM to nSBs communication.

One more link is used to control an internal GBT-SCA to communicate via I2C with the FPGA.

In addition of receiving and transmit slow control command to the central system the main task of the nPDM, as the old one, is to generate and distribute a low jitter synchronous test pulse (that is the reason for its name). These pulses are in a chosen phase relation with the LHC machine clock. The signals are used for a precise timing alignment. This PDM feature was very useful during LHCb Run1 to synchronize in time all the 100,000 channels of the huge the detector. The signals are produced inside the nPDM, using the signal distributed in the LHC ring by the timing and trigger control, a pulse is generated inside the nPDM FPGA and sent to all the nSB and then from them to the front end boards with a precision better than 1 ns.

The pulses signals are acquired from the readout chain of the Muon detector and used from the Off Detector Boards TDCs to measure the time position respect the beam crossing for each channel. This measurement is used to tune any single channel delay register in that way all the pulses reach the Off Detectors readout board at the same time, despite the distance of the channel and the length of the cable.

Each one of the Cardiac delay lines for timing alignment must be calibrated; this is done after the power up of the system sending from the nPDM a 40 MHz clock and changing the electric voltage of the delay line buffer chain to have a total delay of 25 ns.

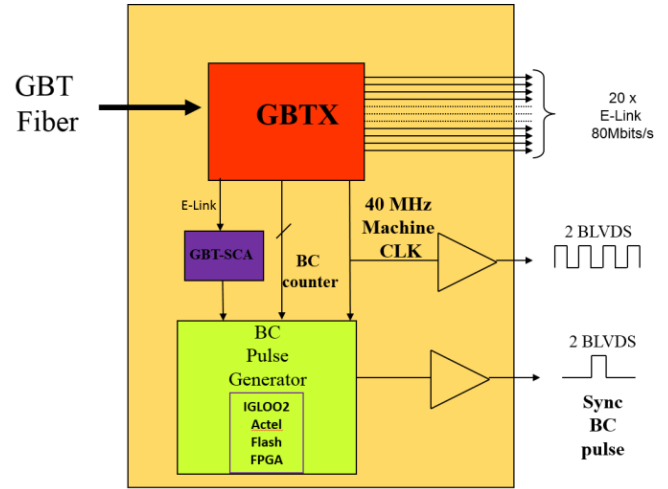

*Fig. 4 new Pulse Distribution Module (nPDM) architecture*

In the architecture of the nPDM, as in the old one, all the functions for timing distribution are performed by an FPGA.

The FPGA must be radiation tolerant and, as in the old version, the choice falls on a FPGA with Flash technology. The Flash memory are order of magnitude more robust of the SRAM to Single Event Upset. The Flash structure of the IGLOO2 from Microsemi corporation (ex Actel) guarantees the flexibility of re-programmability. The architecture and the HDL code of the old PDM FPGA (Actel) can be heavily reused again since it has exactly the same function. As well as for the programmability, the errors due to Single Events Upset must be avoided in the logic. The flip-flops of the logic have the same cross section to SEU as SRAM Cells; in this case, we use, as in the old FPGA, a Triple modular redundancy for registers and State Machines to correct any Single Event Upset

## III. THE NEW SERVICE BOARD (NSB)

The new Service Board (nSB) (Fig. 5) makes possible monitoring and control of Front-End (FE) circuitry, as well as communication between front-end electronics and the external world. The main component of the nSB is a GBT-SCA chip. The GigaBit Transceiver – Slow Control Adapter (GBT–SCA) is used to implement a dedicated control link system for the control and monitoring of the embedded front-end electronics of a High Energy Physics experiment. It communicates with the nPDM GBTx through an point to point 80 Mbps bidirectional e-link. The GBT-SCA provides a number of user-configurable electrical interface ports, able to perform concurrent data transfer operations. The main interface port, used in the nSB, is the block with 16 I2C Masters. During the first design of the LHCb muon ECS we develop a serial protocol between Service Board system and Dialog front end boards mainly based from I2C but using a different physical layer. The Service Board serial link use tree differential LVDS one-way lines: SCL, SDAin, SDout instead of the two bidirectional pull-up lines of I2C: SCL, SDA. Because in the Service Board is the only master there is only one SCL. The SDA lines is used to read and write than must be bidirectional and splitted in two lines.

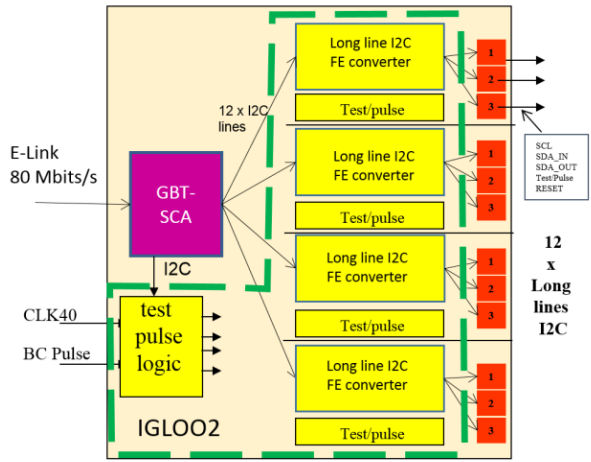

Fig. 5 New Service Board (nSB) internal architecture.

The I2C is normally for short PCB communication. During the design of LHCb we choice to use a customized version with LVDS as physical layer to travel the distance up to 12 meters from the Service Board to the Front-end. We found radiation tolerant commercial LVDS driver chip for the SB side and LVDS IO blocks in the IBM hard radiation Asic technology that used in Dialog. In the old SB the serial protocol, SB-Dialog was totally emulated in the microcontroller. In the nSB we can benefit of the availability of LVDS drivers inside the FPGA and the more resources in the IGLOO2 can be used to realize an on the fly protocol converter from the standard I2C coming from the GBT-SCA to Serial SB interface.

The possibility to use an on the fly converter guaranty no latency from I2C GBT-SCA transaction to SB serial link. Twelve of the sixteen GBT-SCA I2C Master line are than utilized for SB – Front-end communication and one for the GBT-SCA – IGLOO2 communication.

The JTAG master in the GBTx can be used to program the IGLOO2 in the system environment. The new IGLOO2 FPGA as the old FPGA handles the timing signals coming from the nPDM, generates start, and stop signals that commands the Dialog counters in the Front-End Boards. Each one of the 12 serial links connectors have as signals: the serial link interface, the front-end reset and the multifunctional signal Tst/Pulse. The Tst/Pulse can be configured to output timing Pulse, a 40 Mhz clock or start stop signals for front end counter. The Tst/Pls line function depends from Dialog configuration.

## IV. THE BACKPLANE

A special backplane routes the lines from the nPDM to any single nSB (Fig. 6). The twenty 80 Mbits/s E-Links are point to point bidirectional connection using as physical transport layer the SLVS standard[6], any E-Link uses tree differential pairs signals: the clock, data in and data out.

The choice to route the E-LINK in the backplane is solid and feasible and for distance less than 2 m the error rates is $<1 \cdot 10^{-13}$[7]. The 40 MHz clock signals and the pulse signal, distributed from the PDM, instead use BLVDS and can follow the same layout of previous Service Board backplane. The Backplane can hold twenty nSB and the nPDM occupies the central slot to optimize the board layout.

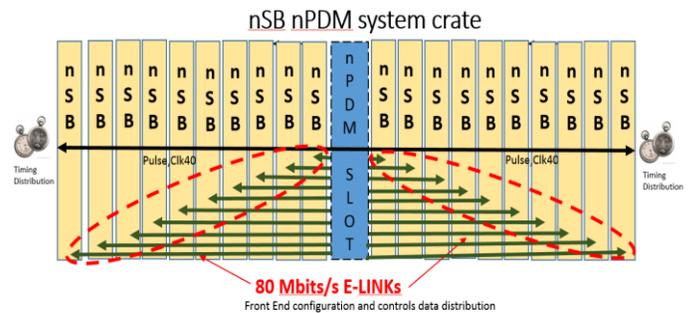

Fig. 6 nSB e nPDM system crate architecure.

## V. CONCLUSIONS

A complete new LHCb muon chamber Front End control system is proposed for LHCb upgrade before run3. The new system is designed to fit exactly in the old system without modify architecture and cabling structure saving previous investments. The new system benefits of the new CERN Radiation Hard Optical Link Project GBT and uses the state of the art of the Flash Field Programmable Gate Array with multiple IO standard capability. The availability of 1.6 Gbits/s links and most powerful computers permits to directly control the system from main computers. Thanks to this high speed data link we use a brutal force data transmission approach instead of the clever idea of intelligent Service Board (simple commands via slow CANbus with intelligent process on board). The architecture of pulse distribution system maintains the totality of the old HDL code reusing it for the new FPGA. Much of the DAQ domain [8] previously written for the old system can be reused.

The use of state of art electronics allows avoiding obsolescence problems for the next ten years.